\documentclass{llncs}
\usepackage[utf8]{inputenc}
\usepackage{graphicx,xspace}
\usepackage{epsfig}
\usepackage{color}
\usepackage{amsmath}
\usepackage{amssymb}
\usepackage[dvipsnames]{xcolor}
\usepackage{pgf}
\usepackage{tikz}
\usetikzlibrary{decorations,arrows}
\usetikzlibrary{decorations.pathmorphing}
\usepackage[british]{babel}

\newcommand\definesymb[1]{%
\expandafter\newcommand\csname #1#1\endcsname{{\ensuremath{\mathbb{#1}}}}%
}

\newcommand\north{\textsf{North}}
\newcommand\south{\textsf{South}}
\newcommand\east{\textsf{East}}
\newcommand\west{\textsf{West}}

\definesymb{Z}
\definesymb{N}
\definesymb{R}

\newcommand\cR{}

\newcommand\lang{\mathcal L}

\newtheorem{thm}{Theorem}[section]

\newtheorem{defn}[thm]{Definition}

\newtheorem{lem}[thm]{Lemma}

\newtheorem{prop}[thm]{Proposition}

\newtheorem{cor}[thm]{Corollary}

\title{Subshifts, Languages and Logic}

\author{Emmanuel Jeandel\thanks{\email{Emmanuel.Jeandel@lif.univ-mrs.fr}}\inst{1}
\and Guillaume Theyssier\thanks{\email{guillaume.theyssier@univ-savoie.fr}}\inst{2}}

\institute{Laboratoire d'informatique fondamentale de Marseille (LIF)\\
  Aix-Marseille Universit\'e, CNRS\\
  39 rue Joliot-Curie, 13\,453 Marseille Cedex 13, France
  \and LAMA (Universit\'e de Savoie, CNRS)\\ Campus
  Scientifique, 73376 Le Bourget-du-lac cedex FRANCE}

\begin{document}

\maketitle

\begin{abstract}
  We study the Monadic Second Order (MSO) Hierarchy over infinite
  pictures, that is tilings. We give a characterization of existential
  MSO in terms of tilings and projections of tilings. Conversely, we
  characterise logic fragments corresponding to various classes of
  infinite pictures (subshifts of finite type, sofic subshifts).
\end{abstract}

\newcommand\pred[2]{P_{\fcolorbox{gray}{#1}{}} (#2)}
\newcommand\tcolor[1]{\fcolorbox{gray}{#1}{}}

\section{Introduction}
\label{sec:intro}
There is a close connection between words and monadic second-order (MSO) logic.
B\"uchi and Elgot proved for finite words that MSO-formulas correspond
exactly to regular languages. This relationship was developed for other
classes of labeled graphs; trees or infinite words enjoy a similar connection. 
See \cite{Thomas,Matz} for a survey of existing results. Colorings of
the entire plane, i.e tilings, represent a natural generalization of
biinfinite words to higher dimensions, and as such enjoy similar properties.
We plan to study in this paper tilings for the point of view of monadic second-order logic.

Tilings and logic have a shared history. The introduction of tilings can be
traced back to Hao Wang \cite{wangpatternrecoII}, who introduced
his celebrated tiles to study the (un)decidability of
the $\forall \exists \forall$ fragment of first order logic.
The undecidability of the domino problem by his PhD Student Berger
\cite{Ber-undecidability-dp} lead then to the undecidability of this fragment
\cite{classicaldecisionproblem}. Seese \cite{Kuske,Seese} used the domino problem
to prove that graphs with a decidable MSO theory have a bounded tree width.
Makowsky\cite{makowsky74,poizat80} used the construction by Robinson \cite{Robinson}
to give the first example of a finitely axiomatizable super-stable theory that is
super-stable. More recently, Oger \cite{oger04} gave generalizations of classical results
on tilings to locally finite relational structures. See the survey
\cite{jac:modth} for more details.

Previously, a finite variant of tilings, called tiling pictures, was studied
\cite{GiamRest2,GiamRest}. Tiling pictures correspond to colorings of a \emph{finite}
region of the plane, this region being bordered by special `\texttt{\#}' symbols.
It is proven for this particular model that language recognized by 
EMSO-formulas correspond exactly to so-called finite tiling systems, 
i.e. projections of finite tilings.

The equivalent of finite tiling systems for infinite pictures are
so-called \emph{sofic subshifts} \cite{Weiss}. A \emph{sofic subshift}
represents intuitively local properties and ensures that every point
of the plane behaves in the same way.  As a consequence, there is no
general way to enforce that some specific color, say $\tcolor{white}$
appears at least once.  Hence, some simple first-order existential
formulas have no equivalent as sofic subshift (and even
subshift). This is where the border of \texttt{\#} for finite pictures
play an important role: Without such a border, results on finite
pictures would also stumble on this issue.

We deal primarily in this article with subshifts. See \cite{Alten}
for other acceptance conditions (what we called subshifts of finite type
correspond to A-acceptance in this paper).

Finally, note that all decision problems in our context are non-trivial : To
decide if a universal first-order formula is satisfiable (the domino
problem, presented earlier) is not recursive.
Worse, it is $\Sigma_1^1$-hard to decide if a tiling of the plane exists
where some given color appears infinitely often \cite{Harel,Alten}. As a
consequence, the satisfiability of MSO-formulas is at least $\Sigma_1^1$-hard.

\section{Symbolic Spaces and Logic}
\label{sec:def}

\subsection{Configurations}
Let ${d\geq 1}$ be a fixed integer and consider the discrete lattice
$\ZZ^d$. For any finite set $Q$, a $Q$-configuration is a function from
$\ZZ^d$ to $Q$. $Q$ may be seen as a set of \emph{colors} or \emph{states}.
An element of $\ZZ^d$ will be called a \emph{cell}. A configuration will
usually be denoted $C,M$ or $N$.

\begin{figure}[!htb]
\[
	\begin{array}{cc}
M & N\\
        \begin{tikzpicture}[scale=.6]
                \clip (0,1) rectangle (6,8);
                \clip[decorate, decoration={zigzag,segment length=9mm}] (0.5,7.5)   -- (4.5,7.5) -- (4.5,1.5)--(0.5,1.5) -- (0.5,7.5);              
                \filldraw[fill=blue]    (0,7) rectangle +(1,1);
                \filldraw[fill=lime]  (1,7) rectangle +(1,1);
                \filldraw[fill=olive]   (2,7) rectangle +(1,1);
                \filldraw[fill=lime]   (3,7) rectangle +(1,1);
                \filldraw[fill=lime]   (4,7) rectangle +(1,1);  
                \filldraw[fill=white] (0,6) rectangle +(1,1);
                \filldraw[fill=blue]    (1,6) rectangle +(1,1);
                \filldraw[fill=white] (2,6) rectangle +(1,1);
                \filldraw[fill=blue]    (3,6) rectangle +(1,1);
                \filldraw[fill=blue]    (4,6) rectangle +(1,1);
                \filldraw[fill=lime]  (1,5) rectangle +(1,1);
                \filldraw[fill=lime]   (2,5) rectangle +(1,1);
                \filldraw[fill=lime]  (3,5) rectangle +(1,1);
                \filldraw[fill=olive]   (4,5) rectangle +(1,1);
                \filldraw[fill=lightgray] (0,5) rectangle +(1,1);
                \filldraw[fill=white] (1,4) rectangle +(1,1);
                \filldraw[fill=blue]    (2,4) rectangle +(1,1);
                \filldraw[fill=blue]    (3,4) rectangle +(1,1);
                \filldraw[fill=white] (0,4) rectangle +(1,1);
                \filldraw[fill=lightgray]   (4,4) rectangle +(1,1);
                \filldraw[fill=olive]   (1,3) rectangle +(1,1);
                \filldraw[fill=olive]   (2,3) rectangle +(1,1);
                \filldraw[fill=lime]  (3,3) rectangle +(1,1);
                \filldraw[fill=lime]  (4,3) rectangle +(1,1);
                \filldraw[fill=lime]  (0,3) rectangle +(1,1);           
                \filldraw[fill=olive]   (1,2) rectangle +(1,1);
                \filldraw[fill=lightgray]   (2,2) rectangle +(1,1);
                \filldraw[fill=white]  (3,2) rectangle +(1,1);
                \filldraw[fill=blue]    (0,2) rectangle +(1,1);
                \filldraw[fill=white]    (4,2) rectangle +(1,1);
                \filldraw[fill=olive]   (1,1) rectangle +(1,1);
                \filldraw[fill=olive]   (2,1) rectangle +(1,1);
                \filldraw[fill=lime]  (3,1) rectangle +(1,1);
                \filldraw[fill=white]  (4,1) rectangle +(1,1);
                \filldraw[fill=blue]    (0,1) rectangle +(1,1);
        \end{tikzpicture}
&
        \begin{tikzpicture}[scale=.6]
                \clip (0,1) rectangle (6,8);
                \clip[decorate, decoration={zigzag,segment length=9mm}] (0.5,7.5)   -- (4.5,7.5) -- (4.5,1.5)--(0.5,1.5) -- (0.5,7.5);
                \filldraw[fill=lime]    (0,7) rectangle +(1,1);
                \filldraw[fill=lime]  (1,7) rectangle +(1,1);
                \filldraw[fill=olive]   (2,7) rectangle +(1,1);
                \filldraw[fill=olive]   (3,7) rectangle +(1,1);
                \filldraw[fill=lime]  (4,7) rectangle +(1,1);   
                \filldraw[fill=white] (0,6) rectangle +(1,1);
                \filldraw[fill=blue]    (1,6) rectangle +(1,1);
                \filldraw[fill=white] (2,6) rectangle +(1,1);
                \filldraw[fill=blue]    (3,6) rectangle +(1,1);
                \filldraw[fill=white]    (4,6) rectangle +(1,1);
                \filldraw[fill=lime]    (0,5) rectangle +(1,1);
                \filldraw[fill=lime]  (1,5) rectangle +(1,1);
                \filldraw[fill=olive]   (2,5) rectangle +(1,1);
                \filldraw[fill=olive]   (3,5) rectangle +(1,1);
                \filldraw[fill=lime]  (4,5) rectangle +(1,1);   
                \filldraw[fill=white] (0,4) rectangle +(1,1);
                \filldraw[fill=blue]    (1,4) rectangle +(1,1);
                \filldraw[fill=white] (2,4) rectangle +(1,1);
                \filldraw[fill=blue]    (3,4) rectangle +(1,1);
                \filldraw[fill=white]    (4,4) rectangle +(1,1);
                \filldraw[fill=lime]    (0,3) rectangle +(1,1);
                \filldraw[fill=lime]  (1,3) rectangle +(1,1);
                \filldraw[fill=olive]   (2,3) rectangle +(1,1);
                \filldraw[fill=olive]   (3,3) rectangle +(1,1);
                \filldraw[fill=lime]  (4,3) rectangle +(1,1);   
                \filldraw[fill=white] (0,2) rectangle +(1,1);
                \filldraw[fill=blue]    (1,2) rectangle +(1,1);
                \filldraw[fill=white] (2,2) rectangle +(1,1);
                \filldraw[fill=blue]    (3,2) rectangle +(1,1);
                \filldraw[fill=white]    (4,2) rectangle +(1,1);
                \filldraw[fill=lime]    (0,1) rectangle +(1,1);
                \filldraw[fill=lime]  (1,1) rectangle +(1,1);
                \filldraw[fill=olive]   (2,1) rectangle +(1,1);
                \filldraw[fill=olive]   (3,1) rectangle +(1,1);
                \filldraw[fill=lime]  (4,1) rectangle +(1,1);
        \end{tikzpicture}
\end{array}
\]
\caption{Two configurations}
\label{conf:example}
\end{figure}
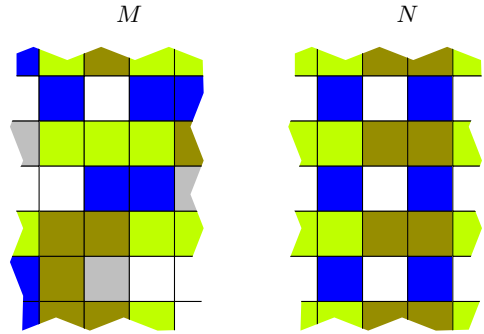

Fig.~\ref{conf:example} shows an example of two different
configurations of $\ZZ^2$ over a set $Q$ of $5$ colors. As a
configuration is infinite, only a finite fragment of the
configurations is represented in the figure. The reader has to use his
imagination to decide what colors do appear in the rest of the
configuration. We choose not to represent which cell of the picture is
the origin $(0,0)$ (we use only translation invariant properties).


\begin{figure}[!htb]
\[
        \begin{tikzpicture}[scale=.6]
                \filldraw[fill=lime]   (3,7) rectangle +(1,1);
                \filldraw[fill=lime]   (4,7) rectangle +(1,1);  
                \filldraw[fill=white] (2,6) rectangle +(1,1);
                \filldraw[fill=blue]    (3,6) rectangle +(1,1);
        \end{tikzpicture}
\]
\caption{A pattern $P$. $P$ appears in $M$ but presumably not in $N$}
\label{pattern:example}
\end{figure}
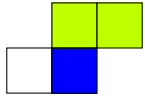

A \emph{pattern} is a partial configuration. A pattern ${P :
  X\rightarrow Q}$ where $X \subseteq \ZZ^2$ occurs in ${C\in Q^{\ZZ^d}}$ at position $z_0$ if
\[\forall z\in X,\ C(z_0+z)=P(z).\]
We say that $P$ occurs in $C$ if it occurs at some position in $C$. As
an example the pattern $P$ of Fig~\ref{pattern:example} occurs in the
configuration $M$ but not in $N$ (or more accurately not on the finite
fragment of $N$ depicted in the figure).  A finite pattern is a
partial configuration of finite domain. All patterns in the following
will be finite. The \emph{language} $\lang(C)$ of a configuration $C$
is the set of finite patterns that occur in $C$. We naturally extend
this notion to sets of configurations.

A \emph{subshift} is a natural concept that captures both the notion of
\emph{uniformity} and \emph{locality}: the only description
``available'' from a configuration $C$ is the finite patterns it
contains, that is $\lang(C)$.  Given a set $\cal F$ of patterns, let
$X_{\cal F}$ be the set of all configurations where no patterns of
$\cal F$ occurs.
\[
	X_{\cal F} = \{ C | \lang(C)\cap \mathcal{F}=\emptyset \}
\]
$\cal F$ is usually called the set of forbidden patterns or the
\emph{forbidden language}. {\cR A set of the form $X_{\cal F}$ is called a
\emph{subshift}.

 A subshift can be equivalentely defined by topology
  considerations. Endow the set of configurations $Q^{\ZZ^d}$ with the
  product topology: A sequence $(C_n)_{n \in \NN}$ of configurations
  converges to a configuration $C$ if the sequence ultimately agree
  with $C$ on every $z \in \ZZ^2$. Then a subshift is a closed subset
  of $Q^{\ZZ^d}$ also closed by shift maps. }

A \emph{subshift of finite type} (or \emph{tiling}) correspond to a finite
set $\cal F$: it is the set of configurations $C$ such that no pattern
in $\cal F$ occurs in $C$. If all patterns of $\cal F$ are of diameter $n$,
this means that we only have to see a configuration through a window of
size $n$ to know if it is a tiling, hence the locality.

Given two state sets $Q_1$ and $Q_2$, a projection is a map ${\pi:
  Q_1\rightarrow Q_2}$. We naturally extend it to ${\pi :
  Q_1^{\ZZ^d}\rightarrow Q_2^{\ZZ^d}}$ by $\pi(C)(z) = \pi(C(z)).$ A
\emph{sofic subshift} of state set $Q_2$ is the image by some
projection $\pi$ of some subshift of finite type of state set
$Q_1$. It is also a subshift {\cR(clearly closed by shift maps, and
  topologically closed because projections are continuous maps on a
  compact space)}. A sofic subshift is a natural object in tiling
theory, although quite never mentioned explicitly. It represents the
concept of \emph{decoration}: some of the tiles we assemble to obtain
the tilings may be decorated, but we forgot the decoration when we
observe the tiling.  \newpage
\subsection{Structures}
\label{sec:struc}

{\cR From now on, we restrict to dimension 2.}  A configuration will
be seen in this article as an infinite structure.  The signature
$\tau$ contains four unary maps $\north$, $\south$, $\east$, $\west$
and a predicate $P_c$ for each color $c \in Q$.

A configuration $M$ will be seen as a structure $\mathfrak{M}$ in the
following way: 
\begin{itemize}
	\item The elements of $\mathfrak{M}$ are the points of $\ZZ^2$.
	\item $\north$ is interpreted by $\north^\mathfrak{M}((x,y)) = (x,y+1)$, $\east$ is
	  interpreted by
	  ${\east^\mathfrak{M}((x,y)) = (x+1,y)}$. $\south^\mathfrak{M}$ and
	  $\west^\mathfrak{M}$ are interpreted similarly
	\item $P^\mathfrak{M}_c((x,y))$ is true if and only if the point at
	  coordinate $(x,y)$ is of color $c$, that is if $M(x,y) = c$.
\end{itemize}

As an example, the configuration $M$ of Fig.~\ref{conf:example} has three
consecutive cells with the color $\tcolor{lime}$. That is, the
following formula is true:
\[
	\mathfrak{M} \models \exists z, \pred{lime}{z} \wedge \pred{lime}{\east(z)}
	\wedge \pred{lime}{\east(\east(z))}
\]

As another example, the following formula states that the configuration has
a vertical period of $2$ (the color in the cell $(x,y)$ is the same as the
color in the cell $(x,y+2)$). The formula is false in the structure $\mathfrak{M}$ and
true in the structure $\mathfrak{N}$ (if the reader chose to color the
cells of $N$ not shown in the picture correctly):

\[	
	\forall z,
\left\{
\begin{array}{c}
	\pred{olive}{z} \implies \pred{olive}{\north(\north(z))}\\
	\pred{white}{z} \implies \pred{white}{\north(\north(z))}\\
	\pred{blue}{z} \implies \pred{blue}{\north(\north(z))}\\
	\pred{lime}{z} \implies \pred{lime}{\north(\north(z))}\\
	\pred{lightgray}{z} \implies \pred{lightgray}{\north(\north(z))}\\
\end{array}
\right.
	\]

\subsection{Monadic Second-Order Logic}
\label{sec:mso}
This paper studies connection between subshifts (seen as structures
as explained above) and monadic second order sentences. First order
variables ($x$, $y$, $z$, ...) are interpreted as points of $\ZZ^2$
and (monadic) second order variables ($X$, $Y$, $Z$, ...) as subsets
of~$\ZZ^2$.

Monadic second order formulas are defined as follows:
\begin{itemize}
\item a term is either a first-order variable or a function ($\south$,
  $\north$, $\east$, $\west$) applied to a term ;
\item atomic formulas are of the form $t_1=t_2$
  or $X(t_1)$ where $t_1$ and $t_2$ are terms and $X$ is either a second order variable or a color predicate ;
\item formulas are build up from atomic formulas by means of boolean
  connectives and quantifiers $\exists$ and $\forall$ (which can be
  applied either to first-order variables or second order variables).
\end{itemize}

A formula is \emph{closed} if no variable occurs free in it. A formula
is FO if no second-order quantifier occurs in it. A formula is EMSO if
it is of the form 
\[\exists X_1,\ldots,\exists X_n, \phi(X)\] where $\phi$ is FO.  Given a
formula $\phi(X_1,\ldots,X_n)$ with no free first-order variable and
having only $X_1,\ldots, X_n$ as free second-order variables, a
configuration $M$ together with subsets ${E_1,\ldots, E_n}$ is a model
of $\phi(X_1,\ldots,X_n)$, denoted
\[(M, E_1,\ldots, E_n) \models \phi(X_1,\ldots,X_n),\] if $\phi$ is
satisfied (in the usual sense) when $M$ is interpreted as $\mathfrak
M$ (see previous section) and $E_i$ interprets $X_i$.

\subsection{Definability}
This paper studies the following problems: Given a formula $\phi$ of
some logic, what can be said of the configurations that satisfy
$\phi$? Conversely, given a subshift, what kind of formula can
characterise it?
\begin{definition}
	A set $S$ of $Q$-configurations is defined by $\phi$ if
	\[ S = \left\{ M \in Q^{\ZZ^2} \middle| \mathfrak{M} \models \phi \right\}\]
	
	Two formulas $\phi$ and $\phi'$ are equivalent iff they define the same
	set of configurations.
	
    A set $S$ is $\cal C$-definable if it is defined by a formula  $\phi \in \cal C$.		
\end{definition}
Note that a definable set is always closed by shift (a shift between $2$ configurations 
induces an isomorphism between corresponding structures).
It is not always closed: The set of $\{\tcolor{lime}, \tcolor{white}\}$-configurations 
  defined by the formula $\phi: {\exists z, \pred{lime}{z}}$ contains all
  configurations except the all-white one, hence is not closed.

  When we are dealing with MSO formulas, the following remark is
  useful: second-order quantifiers may be represented as projection
  operations on sets of configurations. We formalize now this notion.

If $\pi : Q_1 \mapsto Q_2$ is a projection and $S$ is a set of
$Q_1$-configurations, we define the two following operators:
\[
	\begin{array}{rcl}
	E(\pi)(S) &=&\displaystyle \left\{ M \in (Q_2)^{\ZZ^2} \middle|  \exists N \in	  (Q_1)^{\ZZ^2}, \pi(N) = M \wedge N \in S \right\}\\[2ex]
	A(\pi)(S) &=&\displaystyle \left\{ M \in (Q_2)^{\ZZ^2} \middle|  \forall N \in	  (Q_1)^{\ZZ^2}, \pi(N) = M \implies N \in S\right\}
	\end{array}
	\]

Note that $A$ is a dual of $E$, that is $A(\pi)(S) = {^cE}(\pi)({^cS})$ where
$^c$ represents complementation.

\begin{proposition}
	\mbox{}
\label{project}	
\begin{itemize}
	\item
	A set $S$ of $Q$-configurations is EMSO-definable if and only if there
	exists a set $S'$ of $Q'$ configurations and a map $\pi: Q' \mapsto Q$
	such that $S = E(\pi)(S')$ and $S'$ is FO-definable.
	\item
	  The class of MSO-definable sets is the closure of the class of
       FO-definable sets by the operators $E$ and $A$.
	   \end{itemize}
\end{proposition}	
\begin{proof}[Sketch] We prove here only the first item.
\begin{itemize}
	\item Let $\phi = \exists X, \psi$ be a EMSO formula that defines a set $S$ of $Q$-configurations.
	  Let $Q' = Q \times \{ 0,1\}$ and $\pi$ be the canonical projection from $Q'$ to $Q$.

	  Consider the formula $\psi'$ obtained from $\psi$ by replacing $X(t)$
	  by $\vee_{c \in Q} P_{(c,1)}(t)$ and $P_c(t)$ by $P_{(c,0)}(t) \vee P_{(c,1)}(t)$.
	  
      Let $S'$ be a set of $Q'$ configurations defined by $\psi'$. Then is it
	  clear that $S = E(\pi)(S')$. The generalization to more than one
	  existential quantifier is straightforward.
	  
	\item Let $S = E(\pi)(S')$ be a set of $Q$ configurations, and $S'$
	  FO-definable by the formula $\phi$.	  	  
	  Denote by $c_1 \dots c_n$ the elements of $Q'$.
	  Consider the formula $\phi'$ obtained from $\phi$ where each $P_{c_i}$ is 
	  replaced by $X_i$.
      Let
	  
	  \[
\psi = \exists X_1, \dots, \exists X_n,
\left\{
	  \begin{array}{l}
   \forall z, \vee_i X_i(z)  \\
   \forall z, \wedge_{i\not=j} (\neg X_i(z) \vee \neg X_j(z)) \\
   \forall z, \wedge_i \left(X_i{z} \implies P_{\pi(c_i)} (z) \right)\\
   \phi'
   \end{array}\right.
  \]
  
  Then $\psi$ defines $S$.  Note that the formula $\psi$ constructed
  above is of the form ${\exists X_1, \dots, \exists X_n (\forall z, \psi'(z))
    \wedge \phi'}$.  This will be important later.\qed
\end{itemize}
\end{proof}

Second-order quantifications will then be regarded in this paper either as
projections operators or sets quantifiers.

\section{Hanf Locality Lemma and EMSO}

\label{sec:collapse}

The first-order logic has a property that makes it suitable to deal with
tilings and configurations: it is local. This is illustrated by Hanf's
lemma \cite{hanf65,EF:finmt,Libkin}.
\begin{definition}
  Two $Q$-configurations $M$ and $N$ are $(n,k)$-equivalent if for each $Q$-pattern $P$ of size $n$:
  \begin{itemize}
	  \item If $P$ appears in $M$ less than $k$ times, then $P$ appears the
		exact same number of times in $M$ and in $N$
	  \item If $P$ appears in $M$ more than $k$ times, then $P$ appears in
		$N$ more than $k$ times
  \end{itemize}	
\end{definition}
This notion is indeed an equivalence relation. Given $n$ and $k$, it
is clear that there is only finitely many equivalence classes for this
relation.

The Hanf's local lemma can be formulated in our context as follows:
\begin{theorem}
	For every FO formula $\phi$, there exists $(n,k)$ such that
	\begin{center}
	if $M$ and $N$ are $(n,k)$ equivalent, then
	$\mathfrak{M} \models \phi \iff \mathfrak{N} \models \phi$
	\end{center}
\end{theorem}	

\begin{corollary}
	Every FO-definable set is a (finite) union of some $(n,k)$-equivalence classes.
\end{corollary}	
This is theorem 3.3 in  \cite{GiamRest}, stated for finite configurations.
Lemma 3.5 in the same paper gives a proof of Hanf's Local Lemma in our context.

Given $(P, k)$
we consider the set $S_{=k}(P)$ of all configurations such that the pattern
$P$ occurs exactly $k$ times ($k$ may be taken equal to $0$). The set $S_{\geq k}(P)$ is the set of all
configurations such that the pattern $P$ occurs more than $k$ times.

We may rephrase the preceding corollary as:
\begin{corollary}
  \label{fosets}
	Every FO-definable set is a positive combination (i.e. unions and intersections) of some $S_{=k}(P)$ and some $S_{\geq k}(P)$
\end{corollary}	

\begin{theorem}
\label{thm:emso}
Every EMSO-definable set can be defined by a formula $\phi$ of the form:
\begin{align*}
  \exists X_1,\ldots,\exists X_n,\ &\bigl(\forall z_1,\,
  \phi_1(z_1, X_1,\ldots, X_n)\bigr)\\ &\wedge (\exists z_1, \dots,
  \exists z_p,\, \phi_2(z_1 \dots z_p,X_1,\ldots, X_n)\bigr),
\end{align*}
 where
  $\phi_1$ and $\phi_2$ are quantifier-free formulas.
\end{theorem}
See \cite[Corollary 4.1]{Thomas} or  \cite[Corollary 4.2]{Thomas2} for a similar result.
{\cR This result is an easy consequence of \cite[Theorem 3.2]{SchBar} (see
  also the corrigendum). We include here a full proof.}
\begin{proof}
	Let $\cal C$ be the set of such formulas.	
	We proceed in three steps:
	\begin{itemize}
        \item Every EMSO-definable set is the projection of a positive combination of
          some $S_{=k}(P)$ and $S_{\geq k}(P)$ (using
          prop. \ref{project} and the preceding corollary)
		\item Every $S_=(P,k)$ (resp. ${S_{\geq}(P,k)}$) is $\cal C$-definable
		\item $\cal C$-definable sets are closed by (finite) union, intersection  and projections.
	\end{itemize}		
        $\cal C$-definable sets are closed by projection using the
        equivalence of prop. \ref{project} in the two directions, the
        note at the end of the proof and some easy formula
        equivalences. The same goes for intersection.
	
	Now we prove that $\cal C$-definable sets are closed by union. The
	difficulty is to ensure that we use only {\cR one universal quantifier}.
	Let $\phi$ and $\phi'$ be two $\cal C$-formulas defining sets $S_1$ and $S_2$.
	We can suppose that $\phi$ and $\phi'$ use the same numbers of second-order
	quantifiers and of first-order existential quantifiers.
	
Then the formula 
{\cR
\begin{align*}
  \exists X, \exists X_1, \dots, \exists X_n,& \forall z_1, \left\{
    \begin{array}{r}
      X(z_1) \iff X(\north(z_1))\\
      X(z_1) \iff X(\east(z_1))\\
      X(z_1) \implies \phi_1(z_1, X_1 \dots X_n)\\
      \neg X(z_1) \implies \phi'_1(z_1, X_1 \dots X_n)\\
    \end{array}
  \right.\\
  &\wedge \exists z_1, \dots, \exists z_p \bigvee\left.
    \begin{array}{r}
      X(z_1) \wedge \phi_2(z_1 \dots z_p, X_1 \dots X_n)\\
      \neg X(z_1) \wedge \phi'_2(z_1 \dots z_p, X_1 \dots X_n)
    \end{array}
  \right.
\end{align*}
}
defines $S_1 \cup S_2$ (the disjunction is obtained through variable
$X$ which is forced to represent either the empty set or the whole
plane $\ZZ^2$).

It is now sufficient to prove that a $S_{=k}(P)$ set (resp. a $S_{\geq k}(P)$
set) is definable by a $\cal C$-formula.
Let $\phi_P(z)$ be the quantifier-free formula such that $\phi_P(z)$ is true if
and only if $P$ appears at position $z$.

Then $S_{=k}(P)$ is definable by
{\cR 
\begin{align*}
  \exists X_1 \dots \exists X_k \exists A_1, \dots,\exists A_k, &\forall x \left\{
    \begin{array}{l}	
      \wedge_i A_i(x) \iff [A_i(\north(x)) \wedge A_i(\east(x))]\\
      \wedge_i X_i(x) \iff \left[A_i(x) \wedge \neg A_i(\south(x)) \wedge \neg A_i(\west(x))\right]\\
      \wedge_{i \not= j} X_i(x) \implies \neg X_j(x)\\
      (\vee_i X_i(x))  \iff \phi_P(x)\\
    \end{array}		  
  \right.\\
  &\wedge \exists z_1, \dots,\exists z_k, X_1(z_1) \wedge\dots \wedge
  X_k(z_k)
\end{align*}
}
{\cR The formula ensures indeed that $A_i$ represents a quarter of the plane,
$X_i$ being a singleton representing the corner of that plane}. If $k = 0$ this becomes $\forall x, \neg \phi_P(x)$.
To obtain a formula for $S_{\geq k}(P)$, change {\cR the last} $\iff$ to a $\implies$ in the formula.
\qed
\end{proof}


\section{Logic Characterization of SFT and Sofic Subshifts}
\label{sec:sofic}
We start by a characterization of subshifts of finite type (SFTs, i.e tilings).
The problem with SFTs is that they are closed neither by projection nor by
union. As a consequence, the corresponding class of formulas is not
very interesting:

\begin{theorem}
  \label{sft}
	A set of configurations is a SFT if and only if it is defined by a
	formula of the form
    \[\forall z,\, \psi(z)\]
    where $\psi$ is quantifier-free.
\end{theorem}
Note that there is only one quantifier in this formula. Formulas with more
than one universal quantifier do not always correspond to SFT: This is due
to SFTs not being closed by union. 
\begin{proof}
 Let $P_1 \dots P_n$ be patterns. To each $P_i$ we associate the
 quantifier-free formula $\phi_{P_i}(z)$ which is true if and only
 if $P_i $ appears at the position $z$.
 Then the subshifts that forbids patterns $P_1 \dots P_n$ is defined by the
 formula:
 \[ \forall z, \neg \phi_{P_1(z)} \wedge \dots \wedge \neg\phi_{P_n(z)} \]
	 
  Conversely, let $\psi$ be a quantifier-free formula.   
  Each term $t_i$ in $\psi$ is of the form $f_i(z)$ where $f_i$ is some combination of
  the functions $\north, \south, \east$ and $\west$, each $f_i$ thus
  representing somehow some vector $z_i$ ($f_i(z) = z+z_i$).
  Let $Z$ be the collection of all vectors $z_i$ that appear in the formula $\psi$.
  Now the fact that $\psi$ is true at the position $z$ only depends on the
  colors of the configurations in points $(z+z_1), \dots, (z+z_n)$, i.e. on the
  \emph{pattern} of domain $Z$ that occurs at position $z$.
  Let $\cal P$ be the set of patterns of domain $Z$ that makes $\psi$ false.  
  Then the set $S$ defined by $\psi$ is the set of configurations where no
  patterns in $\cal P$ occurs, hence a SFT. \qed
\end{proof}

\begin{theorem}
 \label{thm:sofic}
A set $S$ is a sofic subshift if and only if it is definable by a formula of
the form
\[\exists X_1,\ldots, \exists X_n, \forall z_1, \dots, \forall z_p,\,	\psi(X_1,\ldots,X_n,z_1\dots z_p)\]
where $\psi$ is quantifier-free. {\cR Moreover, any such formula is
  equivalent to a formula of the same form but with a single universal
  quantifier ($p=1$)}.
\end{theorem}

Note that the real difficulty in the proof of this theorem is to treat
the only binary predicate, the equality (=). The reader might try to
find a sofic subshift corresponding to the following formula before
reading the proof:
\[
	  \forall x,y, \left(\pred{lime}{x} \wedge \pred{blue}{\east(y)}\right)     \implies x = y
\]

\begin{proof}

  A sofic subshift being a projection of a SFT, one direction {\cR of
    the first assertion} follows from the previous theorem and
  proposition \ref{project}.

	Let $\cal C$ be the class of formulas of the form:
	\[\exists X_1,\ldots, \exists X_n,  \forall z_{1}, \dots,\forall z_{p},\,\psi(X_1,\ldots,X_n,z_1\dots z_{p_i})\]

    Now we prove by induction on the number $p$ of universal quantifiers 
	that each formula of $\cal C$ is equivalent to a formula with only one
	universal quantifier. There is nothing to prove for $p = 1$.
	
    First, we rewrite the formula in conjunctive normal form:
	\[\exists X_1,\ldots, \exists X_n,  \forall z_{1}, \dots,\forall z_{p},\,\wedge_i \psi_i(X_1,\ldots,X_n,z_1\dots z_{p})\]
		where $\psi_i$ is disjunctive.
	This is equivalent to 
	\[\exists X_1,\ldots, \exists X_n,  \wedge_i \forall z_{1}, \dots,\forall z_{p},\,
		\psi_i(X_1,\ldots,X_n,z_1\dots z_{p}) \equiv \exists X_1,\ldots,\exists X_n,  \wedge_i \eta_i\]		
      Now we treat each $\eta_i$ separately.
      $\psi_i$ is a disjunction of four types of formulas:
\[
	\begin{array}{lclclcl}	
		\bullet P_c(f(x)) &\phantom{aa}& \bullet \neg P_c(f(x))&\phantom{aa}&	\bullet f(x) = y &\phantom{aa}& \bullet f(x) \not= y\\
\end{array}
\]
{\cR because terms are made only of bijective functions (compositions of $\north$, $\south$, $\east$, $\west$).}
We may suppose the last case never happens: $\forall x, y,z f(x)\not=y
\vee \psi(x,y,z)$ is equivalent to $\forall x,z, \psi(x,f(x),z)$.  We
may rewrite
	  \[ \psi_i(z_1 \dots z_{p}) \equiv  \epsilon(z_p) \vee z_p = f(z_{k_1}) \vee \dots \vee z_p = f(z_{k_m}) \vee \theta(z_1 \dots z_{p-1})\]
	  (we forgot the second-order variables to simplify notations)

        We may suppose that no formula is of the form $z_p = z_p$.			 
		Now is the key argument: Suppose that there are strictly more that $m$
		values of $z$ such that $\epsilon(z)$ is false. Then given $z_1 \dots
		z_{p-1}$ we may find a $z_p$ such that the formula ${\epsilon(z_p) \vee (z_p =
		  f(z_{k_1})) \vee \dots \vee (z_p = f(z_{k_m}))}$ is false.
		That is, if there are more than $m$ values of $z$ so that $\epsilon(z)$
		is false, then 
		 \[ \forall z_{1}, \dots,\forall z_{p-1},\, \theta(z_1 \dots z_{p-1})    \]
			 must be true.

        As a consequence, our formula $\eta_i$ is equivalent to the disjunction  of
		the formula
		\[
			\forall z_{1}, \dots,\forall z_{p-1},\, \theta(z_1 \dots z_{p-1})
		\]
and the formula
\[
			\exists S_1, \dots, \exists S_m,
			\left\{\begin{array}{r}
                            \Psi_i\\
			\forall z, \vee_i S_i(z) \iff \neg \epsilon(z)\\
			\forall z_{1}, \dots, \forall z_{p-1},\,   S_1(f(z_{k_1})) \vee \dots			\vee S_m(f(z_{k_m})) \vee \theta(z_1 \dots z_{p-1})\\
			\end{array}
			\right.
		\]
{\cR where $\Psi_i$ express that $S_i$ has at most one element and is defined as follows:}
				  \[
                   {\cR \Psi_i\ \overset{def}{=}\ }\exists A, 
				   \forall x
\left\{
	  \begin{array}{r}
		  A(x) \iff A(\north(x)) \wedge A(\east(x))\\
		  S_i(x) \iff A(x) \wedge \neg A(\south(x)) \wedge \neg A(\west(x))	\\
		  \end{array}
\right.				   
				  \]

Simplifying notations, our formula $\eta_i$ is  equivalent to 
\[
\forall z_{1}, \dots, \forall z_{p-1},\, \theta(z_1 \dots z_{p-1}) \vee
\exists P_1, \dots, \exists P_q 
\forall z_{1}, \dots, \forall z_{p-1},\, \zeta(z_1 \dots z_{p-1})	 \]
which is equivalent to
\[	
	\exists X, \exists P_1,\dots,\exists P_q
	\forall z_1,\dots,\forall z_{p-1}, 
\left\{\begin{array}{r}
		  X(z_1) \iff X(\north(z_1))\\
		  X(z_1) \iff X(\east(z_1))\\
		  X(z_1) \implies \theta(z_1, \dots z_{p-1})\\
		  \neg X(z_1) \implies \zeta(z_1, \dots z_{p-1})\\
	  \end{array}\right.
\]
Now report this new formula instead of $\eta_i$ to obtain
a formula 
\[
\exists X_1,\ldots, \exists X_n,  \wedge_i \exists R_{1}, \dots, \exists R_{q_i}, \forall z_{1},
\dots, \forall z_{p-1},\,\theta_i(z_1\dots z_{p_i}, R_1 \dots R_{q_i})
\]
equivalent to 
\[
\exists X_1,\ldots, \exists X_n, \exists R_{11}, \dots,\exists
R_{kq_k}, \forall z_{1}, \dots, \forall z_{p-1},\,\wedge_i
\theta_i(z_1\dots z_{p_i}, R_{i1} \dots R_{iq_i})
\]
We finally obtain a formula of $\cal C$ with $p-1$ universal
quantifiers, and we may conclude by induction.

To finish the proof, a formula with only one universal quantifier
\[
\exists X_1,\ldots, \exists X_n,  \forall z, \theta(z)
\]
defines indeed a sofic subshift (use the proof of theorem~\ref{sft} to
conclude that this formula defines a projection of a SFT, hence a sofic
subshift)\qed

\end{proof}	

\section{Separation Result}
\label{sec:sep}

Theorems~\ref{thm:emso} and \ref{thm:sofic} above suggest that
EMSO-definable subshifts are not necessarily sofic. We will show in
this section that the set of EMSO-definable subshifts is indeed
strictly larger than the set of sofic subshifts.  The proof is based
on the analysis of the computational complexity of forbidden languages.
It is well-known that sofic subshifts have a recursively
enumerable forbidden language. The following theorem shows that the
forbidden language of an MSO-definable subshift can be arbitrarily
high in the arithmetical hierarchy. 

This is not surprising since
arbitrary Turing computation
can be defined via first order formulas (using tilesets)
and second order quantifiers can be used to simulate quantification of
the arithmetical hierarchy. However, some care must be taken to ensure
that the set of configurations obtained is a subshift.

\begin{theorem}
  Let $E$ be an arithmetical set. Then there is an MSO-definable
  subshift with forbidden language $\mathcal F$ such that $E$ reduces
  to $\mathcal F$ (for many-one reduction).
\end{theorem}
\begin{proof}[sketch]
  Suppose that the complement of $E$ is defined as the set of integers
  $m$ such that:
  \[\exists x_1,\forall x_2,\ldots, {\cR\exists/}\forall x_n, R(m,x_1,\ldots,x_n)\]
  where $R$ is a recursive relation. We first build a formula $\phi$
  defining the set of configurations representing a successful
  computation of $R$ on some input $m, x_1,\ldots, x_n$.  Consider $3$
  colors $c_l$, $c$ and $c_r$ and additional second order variables
  $X_1,\ldots,X_n$ and $S_1,\ldots, S_n$.  The input
  $(m,x_1,\ldots,x_n)$ to the computation is encoded in unary on an
  horizontal segment using colors $c_l$ and $c_r$ and variables $S_i$
  as separators, precisely: first an occurrence of $c_l$ then $m$
  occurrences of $c$, then an occurrence of $c_r$ and, for each
  successive $1\leq i\leq n$, $x_i$ positions in $X_i$ before a
  position of $S_i$.  Let $\phi_1$ be the FO formula expressing the
  following:
  \begin{enumerate}
  \item there is exactly $1$ occurrence of $c_l$ and the same for
    $c_r$ and all $S_i$ are singletons;
  \item starting from an occurrence $c_l$ and going east until
    reaching $S_n$, the only possible successions of states are those
    forming a valid input as explained above.
  \end{enumerate}
  Now, the computation of $R$ on any input encoded as above can be
  simulated via tiling constraints in the usual way. Consider
  sufficiently many new second order variables $Y_1,\ldots,Y_p$ to
  handle the computation and let $\phi_2$ be the FO formula expressing
  that:
  \begin{enumerate}
  \item a valid computation starts at the north of an occurrence of $c_l$;
  \item there is exactly one occurrence of the halting state
    (represented by some $Y_i$) in the whole configuration.
  \end{enumerate}
  We define $\phi$ by:
  \begin{align*}
    \exists X_1,\forall X_2,\ldots,\exists{\cR/\forall} X_n,\exists
    S_1,\ldots,\exists S_n,\exists Y_1,\ldots,\exists Y_p,
    \phi_1\wedge\phi_2.
  \end{align*}
  Finally let $\psi$ be the following FO formula: ${(\forall z, \neg
    P_{c_l})\vee (\forall z, \neg P_{c_r})}$. Let $X$ be the set
  defined by $\phi\vee\psi$. By construction, a finite
  (unidimensional) pattern of the form ${c_l c^m c_r}$ appears in some
  configuration of $X$ if and only if ${m\not\in E}$. Therefore $E$ is
  many-one reducible to the forbidden language of $X$.

  To conclude the proof it is sufficient to check that $X$ is
  closed. To see this, consider a sequence $(C_n)_n$ of configurations
  of $X$ converging to some configuration $C$. $C$ has at most one
  occurrence of $c_l$ and one occurrence of $c_r$. If one of these two
  states does not occur in $C$ then ${C\in X}$ since $\psi$ is
  verified. If, conversely, both $c_l$ and $c_r$ occur (once each)
  then any pattern containing both occurrences also occurs in some
  configuration $C_n$ verifying $\phi$. But $\phi$ is such that any
  modification outside the segment between $c_l$ and $c_r$ in $C_n$
  does not change the fact that $\phi$ is satisfied provided no new
  $c_l$ and $c_r$ colors are added. Therefore $\phi$ is also satisfied
  by $C$ and $C\in X$.\qed
\end{proof}

The theorem gives the claimed separation result for subshifts of EMSO.

\begin{corollary}
  There are EMSO-definable subshifts which are not sofic.
\end{corollary}
\begin{proof}
  In the previous theorem, choose $E$, to be the complement of
  the set of integers $m$ for which there is $x$ such that machine $m$
  halts on empty input in less than $x$ steps. $E$ is not recursively
  enumerable and, using the construction of the proof above, it is
  reducible to the forbidden language of an EMSO-definable
  subshift.\qed
\end{proof}

\section{A Characterization of EMSO}
\label{sec:emso}

EMSO-definable sets are projections of FO-definable sets
(proposition~\ref{project}). Besides, sofic subshifts are projections
of subshifts of finite type (or tilings). Previous results show that
the correspondence sofic$\leftrightarrow$EMSO fails. However, we will
show in this section how EMSO can be characterized through projections
of ``locally checkable'' configurations.

Corollary~\ref{fosets} expresses that FO-definable sets are
essentially captured by counting occurrences of patterns up to some
value. The key idea in the following is that this counting can be
achieved by local checkings (equivalently, by tiling constraints),
provided it is limited to a finite and explicitly delimited
region. This idea was successfully used in~\cite{GiamRest} in the
context of picture languages: pictures are rectangular finite patterns
with a border made explicit using a special state (which occurs all
along the border and nowhere else).  
We will proceed here quite differently. Instead of putting special states on
borders of some rectangular zone, we will simply require that two
special subsets of states $Q_0$ and $Q_1$ are present in the
configuration: we call a \emph{$(Q_0,Q_1)$-marked configuration} any
configuration that contains both a color $q\in Q_0$ and some
color $q'\in Q_1$ somewhere. By extension, given a subshift $\Sigma$ over
$Q$ and two subsets ${Q_0\subseteq Q}$ and ${Q_1\subseteq Q}$, the
\emph{doubly-marked set} $\Sigma_{Q_0,Q_1}$ is the set of
$(Q_0,Q_1)$-marked configurations of $\Sigma$. Finally, a
\emph{doubly-marked set of finite type} is a set $\Sigma_{Q_0,Q_1}$
for some SFT $\Sigma$ and some $Q_0,Q_1$.

\begin{lemma}
  \label{lem:count}
  For any finite pattern $P$ and any $k\geq 0$, $S_{=k}(P)$ is the
  projection of some doubly-marked set of finite type. The same result
  holds for $S_{\geq k}(P)$. 

  Moreover, any positive combination (union and intersection) of
  projections of doubly-marked sets of finite type is also the
  projection of some doubly-marked sets of finite type.
\end{lemma}

\begin{proof}[sketch]
  We consider some base alphabet $Q$, some pattern $P$ and some $k\geq
  0$. We will build a doubly-marked set of finite type over alphabet
  ${Q'=Q\times Q_+}$ and then project back on $Q$.  $Q_+$ is itself a
  product of different layers. The first layer can take values
  $\{0,1,2\}$ and is devoted to the definition of the marker subsets
  $Q_0$ and $Q_1$: a state is in $Q_i$ for ${i\in\{0,1\}}$ if and only
  if its value on the layer is $i$.

  We first show how to convert the apparition in a configuration of
  two marked positions, by $Q_0$ and $Q_1$, into a locally
  identifiable rectangular zone. The zone is defined by two opposite
  corners corresponding to an occurrence of some state of $Q_0$ and
  $Q_1$ respectively. This can be done using only finite type
  constraints as follows.  By adding a new layer of states, one can
  ensure that there is a unique occurrence of a state of $Q_0$ and
  maintain everywhere the following information:
  \begin{enumerate}
  \item $N_{Q_0}(z)\equiv$ the position $z$ is at the north of the (unique)	occurrence
    of a state from $Q_0$,
  \item $E_{Q_0}(z)\equiv$ the position $z$ is at the east of the occurrence
    of a state from $Q_0$.
  \end{enumerate}
  The same can be done for $Q_1$. From that, the membership to the
  rectangular zone is defined at any position $z$ by the following
  predicate (see figure~\ref{fig:lemcount}):
  \[Z(z)\equiv N_{Q_0}(z)\not=N_{Q_1}(z)\wedge
  E_{Q_0}(z)\not=E_{Q_1}(z).\] 

  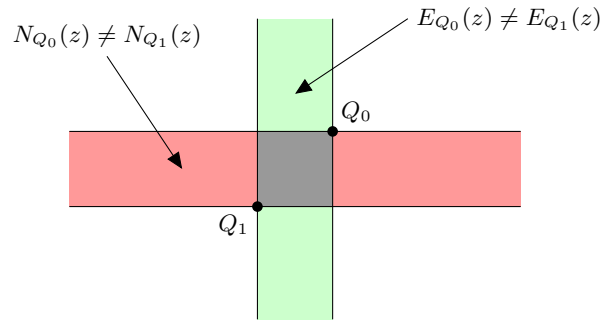
\begin{figure}
    \label{fig:lemcount}
    \centering
    \begin{tikzpicture}
      \fill[fill=green!20] (-.5,-2)--(-.5,2)--(.5,2)--(.5,-2);
      \fill[fill=red!40] (-3,.5)--(3,.5)--(3,-.5)--(-3,-.5);
      \fill[fill=gray!80] (-.5,-.5)--(-.5,.5)--(.5,.5)--(.5,-.5);      
      \draw (-3,.5)--(3,.5);
      \draw (-3,-.5)--(3,-.5);
      \draw (-.5,2)--(-.5,-2);
      \draw (.5,2)--(.5,-2);
      \draw[->,>=triangle 45] (1.5,2) node[right] {$E_{Q_0}(z)\not=E_{Q_1}(z)$} -- (0,1);
      \draw[->,>=triangle 45] (-2.5,1.5) node[above] {$N_{Q_0}(z)\not=N_{Q_1}(z)$} -- (-1.5,0);
      \draw (.5,.5) node[above right] {$Q_0$};
      \draw (-.5,-.5) node[below left] {$Q_1$};
      \fill (.5,.5) circle (2pt);
      \fill (-.5,-.5) circle (2pt);
    \end{tikzpicture}
    \caption{The rectangular zone in dark gray defined by predicate
      $Z(z)$.}
  \end{figure}
  We can also define locally the border of the zone: precisely, cells
  not in the zone but adjacent to it.  Now define $P(z)$ to be true if
  and only if $z$ is the lower-left position in an occurrence of the
  pattern $P$. We add $k$ new layers, each one storing (among other
  things) a predicate $C_i(z)$ verifying
  {\cR \[C_i(z) \Rightarrow Z(z)\wedge P(z) \wedge \bigwedge_{j\not=i}
  \neg C_j(z).\]} Moreover, on each layer $i$, we enforce that exactly $1$
  position $z$ verifies $C_i(z)$: this can be done by maintaining
  north/south and east/west tags (as for $Q_0$ above) and requiring
  that the north (resp. south) border of the rectangular zone sees
  only the north (resp. south) tag and the same for
  east/west. Finally, we add the constraint:
  \[P(z)\wedge Z(z)\Rightarrow \bigvee_{i} C_i\] expressing that each
  occurrence of $P$ in the zone mut be ``marked'' by some $C_i$.
  Hence, the only admissible $(Q_0,Q_1)$-marked configurations are
  those whose rectangular zone contains exactly $k$ occurrences of
  pattern $P$. We thus obtain exactly $S_{\geq k}(P)$ after
  projection. To obtain $S_{= k}(P)$, it suffices to add the
  constraint:
  \[P(z)\Rightarrow Z(z)\] in order to forbid occurrences of $P$
  outside the rectangular zone. 

  To conclude the proof we show that finite unions or intersections of
  projections of doubly-marked sets of finite type are also
  projections of doubly-marked sets of finite type.  Consider two SFT
  $X$ over $Q$ and $Y$ over $Q'$ and two pairs of marker subsets
  $Q_0,Q_1\subseteq Q$ and $Q_0',Q_1'\subseteq Q'$. Let
  ${\pi_1:Q\rightarrow A}$ and ${\pi_2:Q'\rightarrow A}$ be two
  projections.

  First, for the case of union, we can suppose (up to renaming of
  states) that $Q$ and $Q'$ are disjoint and define the SFT $\Sigma$
  over alphabet $Q\cup Q'$ as follows:
  \begin{itemize}
  \item 2 adjacent positions must be both in $Q$ or both in $Q'$;
  \item any pattern forbidden in $X$ or $Y$ is forbidden in $\Sigma$.
  \end{itemize}
  Clearly, ${\pi(\Sigma_{Q_0\cup Q_0',Q_1\cup Q_1'}) =
    \pi_1(X_{Q_0,Q_1})\cup\pi_2(Y_{Q_0',Q_1'})}$ where $\pi(q)$ is
  $\pi_1(q)$ when $q\in Q$ and $\pi_2(q)$ else.
  
  Now, for intersections, consider the SFT $\Sigma$ over
  the fiber product
  
  \[{Q_\times=\{ (q,q') \in Q\times Q' | \pi_1(q) =
	\pi_2(q')\}}\] and defined as follows:
 a pattern is forbidden if its projection on the component
    $Q$ (resp. $Q'$) is forbidden in $X$ (resp. $Y$);

  If we define $\pi$ as $\pi_1$ applied to the $Q$-component of
  states, and if $E$ is the set of configuration of $\Sigma$ such that
  states from $Q_0$ and $Q_1$ appear on the first component and states
  from $Q_0'$ and $Q_1'$ appear on the second one, then we have:
  \[\pi(E) = \pi_1(X_{Q_0,Q_1})\cup\pi_2(Y_{Q_0',Q_1'}).\]
  To conclude the proof, it is sufficient to obtain $E$ as the
  projection of some doubly-marked set of finite type. This can be
  done starting from $\Sigma$ and adding a new component of states
  whose behaviour is to define a zone from two markers (as in the
  first part of this proof) and check that the zone contains
  occurrences of $Q_0$, $Q_1$, $Q_0'$ and $Q_1'$ in the appropriate
  components.\qed
\end{proof}

\begin{theorem}
  A set is EMSO-definable if and only if it is the projection of a
  doubly-marked set of finite type.
\end{theorem}
\begin{proof}
  First, a doubly-marked set of finite type is an FO-definable set
  because SFT are FO-definable (theorem~\ref{sft}) and the restriction
  to doubly-marked configurations can be expressed through a simple
  existential FO
  formula. Thus the projection of a doubly-marked set of finite type
  is EMSO-definable.

  The opposite direction follows immediately from
  proposition~\ref{project} and corollary~\ref{fosets} and the lemma
  above.\qed
\end{proof}

At this point, one could wonder whether considering simply-marked set
of finite type is sufficient to capture EMSO via projections. In fact
the presence of $2$ markers is necessary in the above theorem:
considering the set $\Sigma_{Q_0,Q_1}$ where $\Sigma$ is the full
shift $Q^{\ZZ^2}$ and $Q_0$ and $Q_1$ are distinct singleton subsets
of $Q$, a simple compactness argument allows to show that it is not
the projection of any simply-marked set of finite type.

\section{Open Problems}
\begin{itemize}
\item Is the second order alternation hierarchy strict for MSO
  (considering our model-theoretic equivalence)?
\item One can prove that theorem~\ref{sft} also holds for formulas
  of the form:
    \[\forall X_1 \dots \forall X_n, \forall z,\, \psi(z, X_1\dots X_n)\]
    where $\psi$ is quantifier-free.  Hence, adding universal
    second-order quantifiers does not increase the expression power of
    formulas of theorem~\ref{sft}. More generally, let $\mathcal C$ be the class of formulas
    of the form
  \[\forall X_1,\exists X_2,\ldots, {\cR\forall/}\exists X_n, \forall z_1,\ldots,\forall z_p,\phi(X_1,\ldots,X_n,z_1,\ldots,z_p).\]
  One can check that any formula in $\mathcal C$ defines a
  subshift. Is the second-order quantifiers alternation hierarchy
  strict in $\mathcal C$? On the contrary, do all formulas in $\mathcal C$
  represent sofic subshifts ?
\end{itemize}

\bibliographystyle{plain}
\bibliography{article}

\end{document}